\documentstyle[12pt,epsf,epsfig,wrapfig]{article}
\begin{document}

\begin{center}
{\bfseries Polarization buildup by spin filtering in storage rings}

\vskip 5mm
D.~S.~O'Brien

\vskip 5mm
{\small
{\it
School of Mathematics, Trinity College Dublin, Ireland
}\\

{\it E-mail: donie@maths.tcd.ie
}}
\end{center}

\vskip 5mm
\begin{abstract}
There has been much recent research into polarizing an antiproton beam, instigated by the recent proposal from the PAX (Polarized Antiproton eXperiment) project at GSI Darmstadt \cite{Barone:2005pu}.  It plans to polarize an antiproton beam by repeated interaction with a polarized internal target in a storage ring.  The method of polarization by spin filtering requires many of the beam particles to remain within the ring after scattering off the polarized internal target via electromagnetic and hadronic interactions.  Sets of differential equations which describe the buildup of polarization by spin filtering in many different scenarios have recently been presented and solved [2\,--\,8].
%\cite{Meyer:1994,Milstein:2005bx,Nikolaev:2006gw,O'Brien:2007hu,MacKay:2006,Walcher:2007sj,Buttimore:2007cj}.  
In this paper we add to this literature by investigating a scenario where unpolarized particles are input into the beam at a linearly increasing rate, {\it i.e.}\ the input rate is ramped up. 
\end{abstract}

\vskip 8mm

The spin filtering method of polarization buildup 
%[9\,-11]
\cite{Csonka:1968,Rathmann:1993xf,Rathmann:2004pm} 
consists of a circulating beam repeatedly interacting with a polarized internal target in a storage ring.  Many particles are scattered at small angles but remain in the beam.  This introduces a characteristic acceptance angle $\theta_{\mathrm{acc}}$, scattering above which causes particles to be lost from the beam.  There is also a minimum scattering angle $\theta_{\mathrm{min}}$, corresponding to the Bohr radius of the atoms in the target, below which scattering is prevented by Coulomb screening.  The two physical processes that contribute to polarization buildup in spin filtering are: (a) spin selective scattering out of the ring, and (b) selective spin-flip, {\it i.e.}\ particles in one spin state may be scattered out of the beam (a), or have their spin flipped (b), at a higher rate than particles in the other spin state.  Thus over time one spin state is depleted more than the other leading to a beam polarization.  A problem with this method is that while there is an increase of beam polarization there is a significant decrease in beam intensity, since particles are continuously scattered out of the beam.  We are investigating continuously inputing unpolarized particles into the beam, during spin filtering, to compensate this effect. 

When circulating at frequency $\nu$, for a time $\tau$, in a ring with a polarized internal target of areal density $n$ and polarization ${\mathcal{P}}_{\mathrm{e}}$ oriented normal to the ring plane, (or longitudinally with rotators)
\begin{eqnarray}
\label{eq:HomogeneousSystem}
  \frac{\mathrm{d}}{\mathrm{d}\tau}
\left[
        \begin{array}{c} N \\[2ex] J \end {array}
\right]
\,  = \, - \, n \, \nu
\left[
\begin{array}{ccc}
         I_{\mathrm{\, out}} && {\mathcal{P}}_{\mathrm{e}} \, A_{\mathrm{\, out}}
\\[2ex]
    {\mathcal{P}}_{\mathrm{e}} \, A_{\mathrm{\, all}} -  {\mathcal{P}}_{\mathrm{e}} \, K_{\mathrm{\,in}}
&&
         I_{\mathrm{\, all}} -  D_{\mathrm{\, in}}
\end {array}
\right]
\,
\left[
        \begin{array}{c} N \\[2ex] J \end {array}
\right] \ ,
\end{eqnarray}
 describes the rate of change of the number of beam particles $N(\tau) = N_{\uparrow}(\tau) + N_{\downarrow}(\tau)$ and their total spin $J(\tau) = N_{\uparrow}(\tau) - N_{\downarrow}(\tau)$ \cite{Milstein:2005bx,Nikolaev:2006gw}.  The matrix entries are the spin observables integrated with respect to scattering angle $\theta$ over the following ranges.  The {\bf``in''} subscript refers to particles that are scattered at small angles $\leq \theta_{\mathrm{acc}}$ remaining in the beam, and the {\bf``out''} subscript refers to particles that are scattered out of the beam.  Thus the integrals over scattering angle $\theta$ are labeled {\bf``in''} where the range of integration is $\theta_{\mathrm{min}} \leq \theta \leq \theta_{\mathrm{acc}}$, {\bf``out''} where the range of integration is $\theta_{\mathrm{acc}} < \theta \leq \pi$ and {\bf``all''} $=$ {\bf``in''} $+$ {\bf``out''} where the range of integration is $\theta_{\mathrm{min}} \leq \theta \leq \pi$ as seen in table~1 of ref.~\cite{Buttimore:2007cj}.  $I = {\mathrm{d}}\sigma\,/\,{\mathrm{d}}\Omega$ is the spin averaged differential cross-section and $A$, $K$ and $D$ are the double spin asymmetry, polarization transfer and depolarization spin observables respectively as calculated in ref.~\cite{O'Brien:2006zt}.  The eigenvalues of the above matrix of coefficients are found to be 
\begin{equation}
\lambda_1  \ = \  - \,n\,\nu\, \left(\,I_{\mathrm{\,out}} \,+\, L_{\mathrm{\,in}} \,+\, L_{\mathrm{\,d}}\,\right)
 \hspace*{2em} \mbox{and}  \hspace*{2em}
\lambda_2  \ = \  - \,n\,\nu\, \left(\,I_{\mathrm{\,out}} \,+\, L_{\mathrm{\,in}} \,-\, L_{\mathrm{\,d}}\,\right) \, ,
\end{equation}
where the discriminant $L_{\mathrm{\,d}}$ of the quadratic equation
 for the eigenvalues is
\begin{equation}
   L_{\mathrm{\,d}}
\, =
\, \sqrt{\, {\mathcal{P}}_{\mathrm{e}}^{\,2} \, A_{\mathrm{\,out}} \left( A_{\mathrm{\,all}}
\, -
\, K_{\mathrm{\,in}} \right) \, + \, L_{\mathrm{\,in}}^{\,2} }  \ \, ,
\end{equation}
and
$L_{\mathrm{\,in}} \, = \left( \, I_{\mathrm{\,in}} \, - \, D_{\mathrm{\,in}}\right)\,/\,2$
is a loss of polarization quantity.  Note that $I_{\mathrm{\,out}}$, $L_{\mathrm{\,in}}$ and $L_{\mathrm{\,d}}$ are all positive.  As a consequence the eigenvalues are negative and $\lambda_1 < \lambda_2 < 0$.

 The system above and various alternative scenarios have been developed and solved recently in ref.~\cite{Buttimore:2007cj}.  These scenarios are: 1) spin filtering of a fully stored beam, 2) spin filtering while the beam is being accumulated, {\it i.e.}\ unpolarized particles are continuously being fed into the beam at a constant rate, 3) the particle input rate is equal to the rate at which particles are being lost due to scattering beyond ring acceptance angle, the beam intensity remaining constant, 4) increasing the initial polarization of a stored beam by spin filtering, 5) the input of particles into the beam is stopped after a certain amount of time, but spin filtering continues.  In this paper we add to the literature by investigating a scenario where unpolarized particles are input into the beam at a linearly increasing rate, {\it i.e.}\ the input rate is ramped up.  This is accounted for by the following system of spin evolution equations
\begin{eqnarray}
\label{eq:ODE_for_N}
\frac{d\,N(\tau)}{d\,\tau} & = & \, - \, n \, \nu \ \left[\, I_{{\mathrm{\,out}}} \ N(\tau) \ + \  {{\mathcal{P}}}_{{\mathrm{e}}} \, A_{{\mathrm{\,out}}} \ J(\tau)\, \right] \ + \  \beta \, \tau \,,\\[1ex]
\label{eq:ODE_for_J}
\frac{d\,J(\tau)}{d\,\tau} & = &\, - \, n \, \nu \  \left[\, {{\mathcal{P}}}_{{\mathrm{e}}} \, \left(\,A_{\mathrm{\, all}} -  K_{\mathrm{\,in}}\,\right)\ N(\tau) \  + \ \left(\,I_{\mathrm{\, all}} -  D_{\mathrm{\, in}}\,\right)\ J(\tau)\, \right] \,,
\end{eqnarray}
where $\beta\,\tau$ is the rate at which particles are fed in, the input ramped up at a rate proportional to the time elapsed.  The initial conditions are $N(0) = N_0$ which we may later set to zero, and $J(0) = 0$.  By differentiating eq.(\ref{eq:ODE_for_J}) with respect to $\tau$ and substituting in eq.(\ref{eq:ODE_for_N}) one obtains a second order linear inhomogeneous differential equation for $J(\tau)$:
\begin{eqnarray}
\label{eq:Inhomogeoeous_J_Differential_Equation}
& & \frac{d^{\,2}\,J(\tau)}{d\,\tau^2}  \ - \ \left(\,\lambda_1 + \lambda_2\,\right)\,\frac{d\,J(\tau)}{d\,\tau} \ + \ \lambda_1\,\lambda_2\,J(\tau) \ = \ - \,n\, \nu \,{\mathcal{P}}_{\mathrm{e}}\,\left(\,A_{\mathrm{\,all}} - K_{\mathrm{\,in}}\,\right) \, \beta \, \tau \,,
\,
\end{eqnarray}
the solution of which is
\begin{equation}
\label{eq:Inhomogeneous_J_solution}
J(\tau)  =  F_{\lambda_2\,\lambda_1}\,e^{\,\lambda_1\,\tau} + F_{\lambda_1\,\lambda_2}\,e^{\,\lambda_2\,\tau} + \beta \,\left(\,A_1\,\tau + A_2 \,\right)\,.
\end{equation}
Where for convenience we have defined the constants

\begin{eqnarray}
A_1 \ \equiv \ \frac{-\,n\,\nu\,{\mathcal{P}}_{\mathrm{e}}\,\left(\,A_{\mathrm{\,all}} - K_{\mathrm{\,in}}\,\right)}{\lambda_1\,\lambda_2}
\hspace*{1em} \mbox{and} \hspace*{1em}
A_2 \ \equiv \ \frac{2\,n^{\,2}\,\nu^{\,2}\,{\mathcal{P}}_{\mathrm{e}}\,\left(\,A_{\mathrm{\,all}} - K_{\mathrm{\,in}}\,\right)\left(\,L_{\mathrm{\,in}} + I_{\mathrm{\,out}}\,\right)}{\lambda_1^{\,2}\,\lambda_2^{\,2}} \,,
\end{eqnarray}
\begin{equation}
F_{\lambda_2\,\lambda_1} \ \equiv \ \frac{n\,\nu\,\left(\,A_{\mathrm{\,all}} - K_{\mathrm{\,in}}\,\right)\,N_0\,{\mathcal{P}}_{\mathrm{e}} + \beta \, \left(\,A_1 - \lambda_2\,A_2\,\right)}{\lambda_2 - \lambda_1} \,,
\end{equation}
obtained by imposing the initial conditions $J(0) = 0$ and $N(0) = N_0$ thus $d\,J (0) / d\,\tau = -\,n \, \nu \,{\mathcal{P}}_{\mathrm{e}}\,\left(\,A_{\mathrm{\,all}} - K_{\mathrm{\,in}}\,\right)\,N_0$.  The function $F_{\lambda_1\,\lambda_2}$ is $F_{\lambda_2\,\lambda_1}$ with $\lambda_1$ and $\lambda_2$ interchanged.  
Differentiating eq.(\ref{eq:Inhomogeneous_J_solution}) with respect to $\tau$ and substituting into eq.(\ref{eq:ODE_for_J}) gives an expression for $N(\tau)$:
\begin{eqnarray}
\label{eq:Inhomogeneous_N_Solution}
N(\tau) & = & \frac{-1}{\left(\,A_{\mathrm{\,all}} - K_{\mathrm{\,in}}\,\right)\,{\mathcal{P}}_{\mathrm{e}}}\, \left\{\,F_{\lambda_2\,\lambda_1}\,e^{\,\lambda_1\,\tau}\left(\,L_{\mathrm{\,in}} - L_{\mathrm{\,d}}\,\right) + F_{\lambda_1\,\lambda_2}\,e^{\,\lambda_2\,\tau}\left(\,L_{\mathrm{\,in}} + L_{\mathrm{\,d}}\,\right)\right. \\[1ex]
&   &  \left. \qquad \qquad \qquad \qquad \, + \, \beta \, \left[\,\frac{A_1}{n\,\nu} + \left(\,I_{\mathrm{\,out}} + 2\,L_{\mathrm{\,in}}\,\right)\,\left(\,A_1\,\tau + A_2\,\right)\,\right]
\,\right\} \,.
\end{eqnarray}
As a consistency check it can be seen that the inhomogeneous solutions for $J(\tau)$ and $N(\tau)$ satisfy the initial conditions, and that when $\beta = 0$ they reduce to the solutions of the homogeneous system eq.(\ref{eq:HomogeneousSystem}) presented in refs.~\cite{O'Brien:2007hu,Buttimore:2007cj}.

Dividing $J(\tau)$ by $N(\tau)$ we obtain an expression for the polarization as a function of time ($\tau$),
\begin{equation}
{\mathcal{P}}(\tau) \ = \ \frac{J(\tau)}{N(\tau)} \ = \  \frac{-\,{\mathcal{P}}_{\mathrm{e}}\,\left(\,A_{\mathrm{\,all}} - K_{\mathrm{\,in}}\,\right)}{L_{\mathrm{\,in}} + L_{\mathrm{\,d}}\,\displaystyle{\left[\,\frac{2}{1 - \frac{e^{\,\lambda_1\,\tau}\,F_{\lambda_2\,\lambda_1}\,\left(\,\lambda_2 - \lambda_1\,\right)\, - \, \beta\,\left[\,A_1\,\left(\,1 - \lambda_2\,\tau\,\right) - \lambda_2\,A_2\,\right]
}{e^{\,\lambda_2\,\tau}\,F_{\lambda_1\,\lambda_2}\,\left(\,\lambda_1 - \lambda_2\,\right)\, - \, \beta\,\left[\,A_1\,\left(\,1 - \lambda_1\,\tau\,\right) - \lambda_1\,A_2\,\right]}}\,-\,1\,\right]
}}\,.
\end{equation}
When $\beta = 0$ the above equation simplifies to 
\begin{equation}
{\mathcal{P}}(\tau) \ = \ 
\frac{-\, {\mathcal{P}}_{\mathrm{e}}\,\left(\, A_{\mathrm{\,all}} \, - \, K_{\mathrm{\,in}}\,\right)}{L_{\mathrm{\,in}} \, + \, L_{\mathrm{\,d}} \, \coth\left(L_{\mathrm{\,d}}\, n\,\nu\,\tau\right)}\,,
\end{equation}
which is the solution of the homogeneous case eq.(\ref{eq:HomogeneousSystem}) presented in refs.~\cite{O'Brien:2007hu,Buttimore:2007cj}.

Of interest is the case when $N(0) = N_0 = 0$, {\it i.e.}\ there are no particles in the beam initially.  To obtain this result we set $N_0 = 0$ in the above equation to obtain
\begin{equation}
{\mathcal{P}}(\tau) = \frac{-\,{\mathcal{P}}_{\mathrm{e}}\,\left(\,A_{\mathrm{\,all}} - K_{\mathrm{\,in}}\,\right)}{L_{\mathrm{\,in}} + L_{\mathrm{\,d}}\,\displaystyle{\left[\,\frac{2}{1 - \frac{\left(\,e^{\,\lambda_1\,\tau} - 1\,\right)\,\lambda_2\,A_2 - A_1\,\left(\,e^{\,\lambda_1\,\tau} + \lambda_2\,\tau -1\,\right)}{\left(\,e^{\,\lambda_2\,\tau}-1\,\right)\,\lambda_1\,A_2 - A_1\,\left(\,e^{\,\lambda_2\,\tau} + \lambda_1\,\tau -1\,\right)}}\,-\,1\,\right]
}}\,,
\end{equation}
where for $\beta \neq 0$ the $\beta$ dependence vanishes.  We should note the obvious physical fact that if $N_0 = 0$ and $\beta = 0$, {\it i.e.}\ there are no particles in the beam initially and no particles are fed into the beam, then there will never be any particles in the beam; so measuring the beam polarization is meaningless.  Using a Taylor Series expansion we obtain the approximate initial rate of polarization buildup 
\begin{equation}
   \frac{{\mathrm{d}}\,{\mathcal{P}}}{{\mathrm{d}}\tau} \, \, \approx \, -\, n \, \nu \, {\mathcal{P}}_{\mathrm{e}}
\,
\left( A_{\mathrm{\,all}} \, - \, K_{\mathrm{\,in}} \right)\,,
\end{equation}
identical to that of the homogeneous case eq.(\ref{eq:HomogeneousSystem}) presented in refs.~\cite{O'Brien:2007hu,Buttimore:2007cj}.  The maximum polarization achievable is the limit as time approaches infinity:
\begin{equation}
{\mathcal{P}}_{\mathrm{\,max}} = \lim_{\tau \to \infty} {\mathcal{P}}(\tau) = \frac{-\,{\mathcal{P}}_{\mathrm{e}}\,\left(\,A_{\mathrm{\,all}} - K_{\mathrm{\,in}}\,\right)}{I_{\mathrm{\,all}} - D_{\mathrm{\,in}}} = \frac{-\,{\mathcal{P}}_{\mathrm{e}}\,\left(\,A_{\mathrm{\,all}} - K_{\mathrm{\,in}}\,\right)}{I_{\mathrm{\,out}} + 2\,L_{\mathrm{\,in}}}\,.
\end{equation}
The above expression is only valid for $\beta \neq 0$, the $\beta = 0$ expression is presented in refs.~\cite{O'Brien:2007hu,Buttimore:2007cj}.

The Figure Of Merit (FOM) provides a measure of the quality of the polarized beam, taking into account the trade-off between increasing beam polarization and decreasing beam intensity.  For this inhomogeneous case the FOM is:
\begin{eqnarray}
FOM(\tau) & = & {\mathcal{P}}^{\,2}(\tau)\,N(\tau) = \frac{J^{\,2}(\tau)}{N(\tau)} \ = \  \\[2ex]
& & \hspace*{-7em} \frac{- \,{\mathcal{P}}_{\mathrm{e}}\,\left(\,A_{\mathrm{\,all}} - K_{\mathrm{\,in}}\,\right)\,\left[\,F_{\lambda_2\,\lambda_1}\,e^{\,\lambda_1\,\tau} + F_{\lambda_1\,\lambda_2}\,e^{\,\lambda_2\,\tau} + \beta \,\left(\,A_1\,\tau + A_2 \,\right)\,\right]^{\,2}}{F_{\lambda_2\,\lambda_1}\,e^{\,\lambda_1\,\tau}\left(\,L_{\mathrm{\,in}} + L_{\mathrm{\,d}}\,\right)- F_{\lambda_1\,\lambda_2}\,e^{\,\lambda_2\,\tau}\left(\,L_{\mathrm{\,in}} + L_{\mathrm{\,d}}\,\right) +  \beta \, \left[\,\frac{A_1}{n\,\nu} + \left(\,I_{\mathrm{\,all}} - D_{\mathrm{\,in}}\,\right)\,\left(\,A_1\,\tau + A_2\,\right)\,\right]} \nonumber \,.
\end{eqnarray}
If the particle accumulation rate $\beta\,\tau$ is high enough to make the beam intensity constant or increase with time the FOM will be a monotonically increasing function of time, {\it i.e.}\ it will not have a finite maximum.

\vskip 3mm

This research is funded by the Irish Research Council for Science, Engineering and Technology (IRCSET).

\end{document}